\documentclass[Physsubmission, Phys]{SciPost}

\hypersetup{
    colorlinks,
    linkcolor={red!50!black},
    citecolor={blue!50!black},
    urlcolor={blue!80!black}
}

\usepackage[bitstream-charter]{mathdesign}
\DeclareSymbolFont{usualmathcal}{OMS}{cmsy}{m}{n}
\DeclareSymbolFontAlphabet{\mathcal}{usualmathcal}

\usepackage{graphicx}
\usepackage{amsmath}
\usepackage{amsfonts}
\usepackage[compat=1.1.0]{tikz-feynhand}
\usepackage{float}

\DeclareMathOperator{\sinc}{sinc}
\DeclareMathOperator{\im}{Im}

\newcommand{\hi}{0.5}
\newcommand{\wi}{0.25}
\newcommand{\wid}{0.55}

\begin{document}

\begin{center}{\Large \textbf{
NLO finite system size corrections to $2\to2$ scattering in $\phi^4$ theory using newly derived sum of sinc functions\\
}}\end{center}

\begin{center}
J.F. Du Plessis\textsuperscript{1$\star$} and
W. A. Horowitz\textsuperscript{2}
\end{center}

\begin{center}
{\bf 1} Department of Physics, Stellenbosch University, Private Bag X1, Matieland 7602, South Africa
\\
{\bf 2} Department of Physics, University of Cape Town, Private Bag X3, Rondebosch 7701, South Africa
\\
* 23787295@sun.ac.za
\end{center}

\begin{center}
\today
\end{center}


\definecolor{palegray}{gray}{0.95}
\begin{center}
\colorbox{palegray}{
  \begin{tabular}{rr}
  \begin{minipage}{0.1\textwidth}
    \includegraphics[width=23mm]{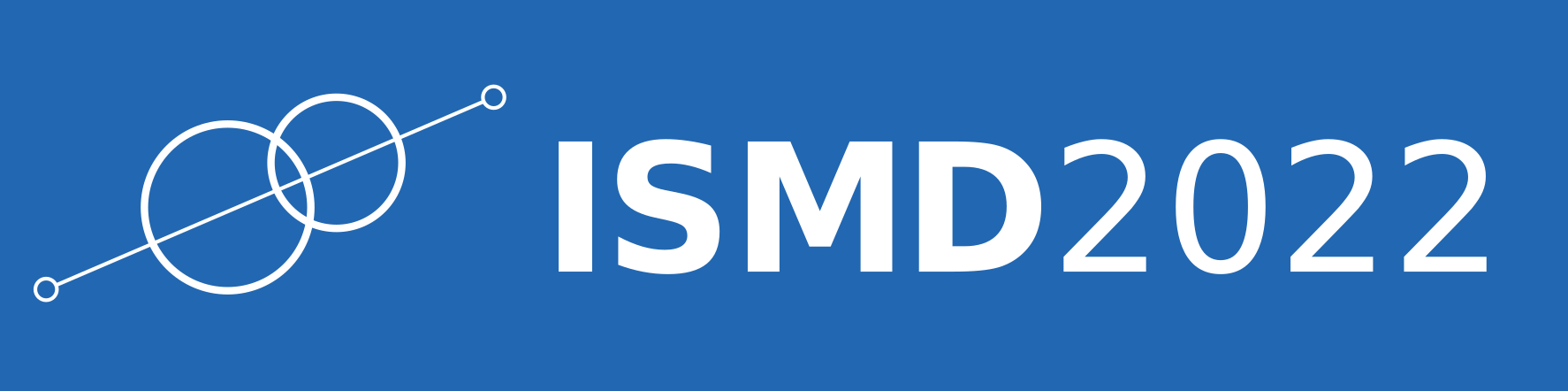}
  \end{minipage}
  &
  \begin{minipage}{0.8\textwidth}
    \begin{center}
    {\it 51st International Symposium on Multiparticle Dynamics (ISMD2022)}\\ 
    {\it Pitlochry, Scottish Highlands, 1-5 August 2022} \\
    \doi{10.21468/SciPostPhysProc.?}\\
    \end{center}
  \end{minipage}
\end{tabular}
}
\end{center}

\section*{Abstract}
{\bf
Previously an equation of state for the relativistic hydrodynamics encountered in heavy-ion collisions at the LHC and RHIC has been calculated using lattice gauge theory methods. This leads to a prediction of very low viscosity, due to the calculated trace anomaly. Finite system corrections to this trace anomaly could challenge this calculation, since the lattice calculation was done in an effectively infinite system. In order to verify this trace anomaly it is sensible to add phenomenologically relevant finite system corrections. We investigate massive $\phi^4$ theory with periodic boundary conditions on $n$ of the 3 spatial dimensions. $2\to2$ NLO scattering is then computed. Using a newly derived formula for an arbitrary dimensional sum of sinc functions, we show that the NLO finite size corrections preserve unitarity.
}


\section{Introduction}
\label{sec:intro}
There is an apparent formation of Quark Gluon Plasma (QGP) in heavy ion collisions \cite{ALICE:2008ngc,ALICE:2010suc,WELLER2017351}, where the correlations between the outgoing low-momentum particles appear to be well described by nearly inviscid relativistic hydrodynamics. This calculation uses an Equation of State (EoS) provided by a lattice QCD calculation that is extrapolated to infinite system size\cite{PhysRevC.87.064906}.

It is currently unclear what happens in QCD just above the transition temperature $T = 180$ MeV. There is strong evidence of a second order phase transition, but the nature of the new phase is largely unknown. It is therefore necessary to understand how reliably the 
 experimental behaviour found in the finite systems (such as heavy ion or parton collisions) can be extrapolated to effectively infinite systems, such as the QGP found in the $\sim0.000001$ seconds after Big Bang. 

A possibly significant assumption to be investigated is that heavy ion collisions can be well approximated as infinite sized systems \cite{Mogliacci:2018oea}. Indeed quenched lattice QCD calculations have shown significant possible corrections dependent on the size of the system \cite{Kitazawa:2019otp}. An analytic derivation of the finite size effects on the equation of state (or equivalently the trace anomaly) is therefore sought. This work is a step in that direction, with the intention to develop and understand the mathematical techniques necessary for a full treatment necessary for finite temperature finite sized QCD.

\section{Finite Sized $\phi^4$ Theory}
Let us consider the $\phi^4$ Lagrangian
\begin{equation}
    \mathcal{L}=\frac12 \partial^\mu\phi\partial_\mu\phi-\frac12 m^2\phi^2-\frac{\lambda}{4!}\phi^4
\end{equation}
in a system with periodic boundary conditions. If we consider $n$ compact spatial dimensions, with the $i$th dimension being parameterized by $[-\pi L_i,\pi L_i]$ with periodic boundary conditions. This discretizes the possible spatial momenta to $\vec{p}=(\frac{k_1}{L_1},\frac{k_2}{L_2},\ldots,\frac{k_n}{L_n},)$ where $\vec{k}\in\mathbb{Z}^n$. In analogy with \cite{Peskin:1995ev} we can define $
    -i\lambda^2 V(p^2)
    \equiv
    \begin{tikzpicture}[baseline=(c.base)]
        \begin{feynhand}
            \setlength{\feynhandarrowsize}{4pt}
            \setlength{\feynhanddotsize}{0mm}
            \vertex (a) at (-\wid,\hi) {};
            \vertex (b) at (-\wid,-\hi) {};
            \vertex [dot] (c) at (-\wi,0) {};
            \vertex [dot] (d) at (\wi,0) {};
            \vertex (e) at (\wid,\hi) {};
            \vertex (f) at (\wid,-\hi) {};
            \draw (a) to (c);
            \draw (b) to (c);
            \draw (c) to [out = 65, in = 115, looseness = 1.75] (d);
            \draw (c) to [out = 295, in = 245, looseness = 1.75] (d);
            \draw (d) to (e);
            \draw (d) to (f);
        \end{feynhand}
    \end{tikzpicture}\nonumber
$ with $p$ being the total incoming momentum. One then finds\cite{Horowitz:2022rpp} in $n=3$ spatial dimensions that, up to NLO, one gets the renormalized
\begin{multline}
    \overline{V}(p^2,\{L_i\}) = -\frac{1}{2(4\pi)^2}\int_0^1 dx\Bigg\{\log\left(\frac{\mu^2}{\Delta^2}\right)\\\cr+ 2\sideset{}{'}\sum_{\vec m\in\mathbb Z^3}\cos(2\pi x \sum m_i p^i L_i) K_0\left(2\pi\sqrt{\Delta^2}\sqrt{\sum (m_i L_i)^2}\right) \Bigg\}.
\end{multline}
Here we recognize the $\log$ term as corresponding to the standard result in infinite $\phi^4$ systems\cite{Peskin:1995ev}. As one would then expect the second term in the integral vanishes in the limit as all $L_i\to\infty$ since $\lim_{x\to\infty}K_0(x)=0$. We can see that we could reduce the effective number of finite dimensions by quite simply taking the corresponding $L_i\to\infty$, since only terms in the sum with the corresponding $m_i=0$ will survive the limit. We then find that, as in the infinite system case,
\begin{equation}
    \mathcal{M}=\lambda\left[1+\lambda\left(\overline{V}(s)+\overline{V}(t)+\overline{V}(u)\right)\right]
\end{equation}
up to NLO, with $s,t$ and $u$ being the usual Mandelstam variables.

\section{Unitarity}
In order to verify that unitarity has stayed intact we will show that the optical theorem holds, no matter how many dimensions $m$ are of finite size. For the optical theorem to hold, we need that
\begin{equation}
    2\im [\mathcal{M}]=\sigma_\text{tot}.
\end{equation}
It is a straight-forward calculation to find
\begin{equation}\label{eq:totcs}
    \sigma_{tot} = \frac{\lambda^2}{16\pi}\frac{\pi^\frac{1-m}{2}}{\Gamma\left(\frac{3-m}{2}\right) }\frac{1}{L\sqrt{s}} \,\sideset{}{^*}\sum_{0\leq l<R^2} \frac{r_m(l)}{\sqrt{R^2-l}^{m-1}}.
\end{equation}
By following \cite{Horowitz:2022rpp} one gets that
\begin{equation}
    2\im [\mathcal{M}]=\frac{\lambda^2}{16\pi}\frac{2R}{L\sqrt{s}}\sum_{\vec{k}\in \mathbb{Z}^m} \sinc\left(2\pi R \|\vec{k}\|\right)
\end{equation}
up to NLO. We can then use \ref{eq:sincSum} to get
\begin{equation}\label{eq:impart}
    2\im [\mathcal{M}]=\frac{\lambda^2}{16\pi}\frac{2R}{L\sqrt{s}}\frac{1}{2 R \pi^\frac{m-1}{2} \,\Gamma\left(\frac{3-m}{2}\right)}\,\sideset{}{^*}\sum_{0\leq l<R^2}\frac{r_m(l)}{\sqrt{R^2-l}^{m-1}}.
\end{equation}
The equivalence of Equations \ref{eq:totcs} and \ref{eq:impart} shows that the Optical Theorem, and therefore Unitarity, holds independent of the amount of compact dimensions.

\section{Conclusion}
By passing all considered self-consistency checks, namely having the correct infinite limit and preserving unitarity, we have shown the viability of the mathematical tools developed and employed. Notably it greatly supports the generalization of the number theoretic formula derived in \ref{sp:sinc}, which has potential implications in number theory.

\section*{Acknowledgements}
W.A.H. wishes to thank the South African National Research Foundation and the SA-CERN Collaboration for support and New Mexico State University for its hospitality, where part of the work was completed. The authors wish to thank Matt Sievert, Alexander Rothkopf, Bowen Xiao, Stan Brodsky, Andrea Shindler, Kevin Bassler and Herbert Weigel for valuable discussions.


\paragraph{Funding information}
The authors wish to thank the South African National Research Foundation and the SA-CERN Collaboration for support.

\begin{appendix}

\section{Sum of Sinc functions}\label{sp:sinc}
We will need a generalization of a formula for the multi-dimensional sum of $\sinc$ functions, namely sums of the form
\begin{equation}
\sum_{\vec{k}\in \mathbb{Z}^m} \sinc(2\pi R \|\vec{k}\|)
\end{equation}
where we define $\sinc(x)=\begin{cases}\frac{\sin(x)}{x} & x\neq0\\1 & x=0\end{cases}$.\\
Since the sum only depends on the magnitude of $\vec{k}$, we can simplify the sum using the \textit{Sum of Squares function} $r_d(n)$ which gives the amount of $\vec{k}\in\mathbb{Z}^d$ with $\|\vec{k}\|^2=n$. Allowing us to use
\begin{equation}\label{eq:Poisson}\sum_{l=0}^\infty r_d(l)f(\sqrt{l})=\sum_{l=0}^\infty r_d(l)\hat{f}(\sqrt{l})\end{equation}
from \cite{JohnsonMcDaniel2012ADC}. To employ their analytically continued version of the Poisson summation formula we can first set $f(r)=\sinc(2\pi R r)$. Then we can calculate $\hat{f}$
\begin{align}
    \hat{f}(p)&=\frac{2\pi^\frac{m}{2}}{\Gamma(\frac{m}{2})}\int_0^\infty \sinc(2\pi R r)\, {}_0 F_1(\frac{m}{2};-\pi^2 p^2 r^2)\,r^{m-1} dr\\
    &=\frac{1}{2 R \pi^\frac{m-1}{2} \,\Gamma(\frac{3-m}{2})}(R^2-p^2)^\frac{1-m}{2}\theta(R^2-p^2)
\end{align}
Now we can apply \ref{eq:Poisson}
\begin{align}
    \sum_{\vec{k}\in \mathbb{Z}^m} \sinc(2\pi R \|\vec{k}\|)&=\sum_{l=0}^\infty r_m(l)\sinc(2\pi R \sqrt{r})\\
    &=\sum_{l=0}^\infty r_m(l)\frac{(R^2-l)^\frac{1-m}{2}}{2 R \pi^\frac{m-1}{2} \,\Gamma(\frac{3-m}{2})}\theta(R^2-l).
\end{align}
Here we should note that if $r_m(R^2)\neq0$ we will get a term with $\theta(0)=\frac12$ which we note will give a $\frac10$ term in the sum for $m>1$. Therefore
\begin{equation}\label{eq:sincSum}
    \sum_{\vec{k}\in \mathbb{Z}^m} \sinc(2\pi R \|\vec{k}\|)=\frac{1}{2 R \pi^\frac{m-1}{2} \,\Gamma(\frac{3-m}{2})}\,\sideset{}{^*}\sum_{0\leq l<R^2}\frac{r_m(l)}{\sqrt{R^2-l}^{m-1}},
\end{equation}
where $\sideset{}{^*}\sum_{0\leq l<R^2}$ means that if $R^2\in\mathbb{Z}$, then its term in the sum has weight $\frac12$. 
The $m=2$ case corresponds directly to a formula of Ramanujan \cite{Hardy1978RamanujanTL}, making this result a generalization thereof.

\end{appendix}



\bibliography{refs.bib}

\begin{thebibliography}{10}
\providecommand{\url}[1]{\texttt{#1}}
\providecommand{\urlprefix}{URL }
\expandafter\ifx\csname urlstyle\endcsname\relax
  \providecommand{\doi}[1]{doi:\discretionary{}{}{}#1}\else
  \providecommand{\doi}{doi:\discretionary{}{}{}\begingroup
  \urlstyle{rm}\Url}\fi
\providecommand{\eprint}[2][]{\url{#2}}

\bibitem{ALICE:2008ngc}
K.~Aamodt \emph{et~al.},
\newblock \emph{{The ALICE experiment at the CERN LHC}},
\newblock JINST \textbf{3}, S08002 (2008),
\newblock \doi{10.1088/1748-0221/3/08/S08002}.

\bibitem{ALICE:2010suc}
K.~Aamodt \emph{et~al.},
\newblock \emph{{Elliptic flow of charged particles in Pb-Pb collisions at 2.76
  TeV}},
\newblock Phys. Rev. Lett. \textbf{105}, 252302 (2010),
\newblock \doi{10.1103/PhysRevLett.105.252302},
\newblock \eprint{1011.3914}.

\bibitem{WELLER2017351}
R.~D. Weller and P.~Romatschke,
\newblock \emph{One fluid to rule them all: Viscous hydrodynamic description of
  event-by-event central p+p, p+pb and pb+pb collisions at s=5.02 tev},
\newblock Physics Letters B \textbf{774}, 351 (2017),
\newblock \doi{https://doi.org/10.1016/j.physletb.2017.09.077}.

\bibitem{PhysRevC.87.064906}
A.~Bzdak, B.~Schenke, P.~Tribedy and R.~Venugopalan,
\newblock \emph{Initial-state geometry and the role of hydrodynamics in
  proton-proton, proton-nucleus, and deuteron-nucleus collisions},
\newblock Phys. Rev. C \textbf{87}, 064906 (2013),
\newblock \doi{10.1103/PhysRevC.87.064906}.

\bibitem{Mogliacci:2018oea}
S.~Mogliacci, I.~Kolb\'e and W.~A. Horowitz,
\newblock \emph{{Geometrically confined thermal field theory: Finite size
  corrections and phase transitions}},
\newblock Phys. Rev. D \textbf{102}(11), 116017 (2020),
\newblock \doi{10.1103/PhysRevD.102.116017},
\newblock \eprint{1807.07871}.

\bibitem{Kitazawa:2019otp}
M.~Kitazawa, S.~Mogliacci, I.~Kolb\'e and W.~A. Horowitz,
\newblock \emph{{Anisotropic pressure induced by finite-size effects in SU(3)
  Yang-Mills theory}},
\newblock Phys. Rev. D \textbf{99}(9), 094507 (2019),
\newblock \doi{10.1103/PhysRevD.99.094507},
\newblock \eprint{1904.00241}.

\bibitem{Peskin:1995ev}
M.~E. Peskin and D.~V. Schroeder,
\newblock \emph{{An Introduction to quantum field theory}},
\newblock Addison-Wesley, Reading, USA,
\newblock ISBN 978-0-201-50397-5 (1995).

\bibitem{Horowitz:2022rpp}
W.~A. Horowitz and J.~F. Du~Plessis,
\newblock \emph{{Finite system size correction to NLO scattering in
  \ensuremath{\phi}4 theory}},
\newblock Phys. Rev. D \textbf{105}(9), L091901 (2022),
\newblock \doi{10.1103/PhysRevD.105.L091901},
\newblock \eprint{2203.01259}.

\bibitem{JohnsonMcDaniel2012ADC}
N.~K. Johnson-McDaniel,
\newblock \emph{A dimensionally continued poisson summation formula},
\newblock Journal of Fourier Analysis and Applications \textbf{18}, 367 (2012).

\bibitem{Hardy1978RamanujanTL}
G.~H. Hardy,
\newblock \emph{Ramanujan: Twelve Lectures on Subjects Suggested by His Life
  and Work},
\newblock Cambridge University Press (1940).

\end{thebibliography}

\nolinenumbers

\end{document}